\documentclass[prb,twocolumn,superscriptaddress,noshowpacs]{revtex4}
\usepackage{graphicx}
\usepackage{physics}
\usepackage{bm}
\usepackage[utf8]{inputenc}
\usepackage{ulem}
\usepackage{color}

\usepackage{esvect}

\begin{document}

\title{Analogue black hole merger in a polariton condensate}

\author{D.~D.~Solnyshkov}
\email{dmitry.solnyshkov@uca.fr}
\affiliation{Institut Pascal, PHOTON-N2, Universit\'e Clermont Auvergne, CNRS, Clermont INP, F-63000 Clermont-Ferrand, France.}
\affiliation{Institut Universitaire de France (IUF), F-75231 Paris, France}

\author{V. Paquelier}
\affiliation{Institut Pascal, PHOTON-N2, Universit\'e Clermont Auvergne, CNRS, Clermont INP, F-63000 Clermont-Ferrand, France.}

\author{C. Balmisse}
\affiliation{Institut Pascal, PHOTON-N2, Universit\'e Clermont Auvergne, CNRS, Clermont INP, F-63000 Clermont-Ferrand, France.}

\author{G.~Malpuech}
\email{guillaume.malpuech@uca.fr}
\affiliation{Institut Pascal, PHOTON-N2, Universit\'e Clermont Auvergne, CNRS, Clermont INP, F-63000 Clermont-Ferrand, France.}

\begin{abstract}
Analogue studies represent an important tool in modern Physics. In particular, analogue gravity had a strong success in the recent years with the demonstrations of Hawking radiation and superradiance of analogue black holes in classical and quantum fluids. So far, the metric of the analogue black holes was mostly fixed by the conditions of the experiment, preventing the simulation of any significant evolution of their properties, such as the change of their mass, their spatial motion, gravitation attraction to other bodies, and, ultimately, black hole mergers. Polariton condensates represent a perfect setting for the analogue simulation of black hole evolution and mergers because of the velocity-dependent losses creating a convergent flow associated with each quantum vortex, which thus becomes an analogue black hole capable of spatial motion. We show that while two vortices are unable to form a common horizon, four or more vortices can exhibit a complete black hole merger, with the radius of the common horizon given by a simple geometrical law. We also discuss the difference between the horizon and the apparent horizon in these analogue black holes with quantized constituents.
\end{abstract}

\maketitle 

\section{Introduction}
Analogue physics demonstrates a strong progress in the recent years~\cite{zohar2015quantum,Eckel2018,viermann2022quantum,braunstein2023analogue,daley2023twenty}. Analogue experiments are useful to simulate not only inaccessible systems, but also systems untractable with modern numerical tools~\cite{Bloch2012,gross2017quantum}. This includes simulations of the early universe inflation and reheating~\cite{Fischer2004,zache2017inflationary,chatrchyan2021analog}, false vacuum decay~\cite{zenesini2024false} -- an example of a non-perturbative many-body problem beyond the limits of theoretical simulations, and analogue black holes~\cite{Barcelo2018}. It was the idea to test the Hawking radiation experimentally~\cite{Unruh1981} that sparkled the field of analogue gravity~\cite{Barcelo2005}. The temperature of the Hawking radiation~\cite{Hawking1974}, determined by the mass of the black holes, is so low for astrophysical black holes that it is undetectable in presence of the cosmic microwave background which has a much higher temperature. Yet, it was important to test the predictions of a quantum field theory in a curved spacetime, and this is where the analogue systems have allowed to demonstrate the validity and even the universality of the Hawking radiation experimentally~\cite{rousseaux2008observation,Weinfurtner2011,Steinhauer2014,steinhauer2016observation,Euve2016,munoz2019observation}.

During the last 40 years, analogue gravity has demonstrated a lot of achievements, including stimulated~\cite{rousseaux2008observation,Weinfurtner2011} and spontaneous Hawking radiation~\cite{steinhauer2016observation} in classical~\cite{bossard2024art} and quantum fluids~\cite{kolobov2021observation}, superradiance~\cite{Torres2017,Braidotti2020,Braidotti2022}, and the ringdown of black hole quasinormal modes~\cite{torres2020quasinormal}. A Kerr black hole with quantized rotation was created recently in liquid helium~\cite{vsvanvcara2024rotating}, but the accessible quantum numbers of the angular momentum were always rather high, making the system close to the classical limit. In analogue black holes, the gravitational attraction is controlled by the convergent flow of the fluid~\cite{Visser1998}, while the rotation of the Kerr black holes is reproduced by the fluid's rotation~\cite{Visser2005}. Both ingredients together give rise to what is called a "draining bathtub configuration". However, similar to a whirl in a bathtub, these analogue black holes have their key properties, such as the mass or the radius of the horizon and the angular momentum, as well as their position, fixed by the conditions of the experiment. This allows studying only relatively weak perturbations, which is perfect for Hawking radiation studies, but appears as a strong limitation if one wishes to simulate the effects linked with the changes of the properties of the black holes, such as the Penrose effect~\cite{Penrose1971}, where the angular momentum of the black hole decreases, or the black hole merger~\cite{buonanno1999effective,LIGO2016}, where all properties of the black holes (position, velocity, angular momentum, mass) need to change.

In recent works~\cite{Solnyshkov2011,Gerace2012,HaiSon2015,solnyshkov2019quantum,jacquet2020polariton,jacquet2022analogue,solnyshkov2024towards,falque2025polariton}, 
it was suggested and shown theoretically and experimentally that polariton condensates represent a very promising setting for analogue gravity experiments. Exciton-polaritons (or simply polaritons) are formed from the strong coupling of excitons and photons, usually achieved in microcavities~\cite{Microcavities}. They combine the strong interactions provided by the excitonic fraction with the fast motion provided by the photonic fraction. The photons also make possible the use of the well-developed optical tools for wavefunction engineering and tomography~\cite{gianfrate2020measurement}, allowing one to measure the density and the velocity of the flow with high precision~\cite{lagoudakis2008quantized}. This has allowed the observation of a black hole horizon~\cite{HaiSon2015} and the studies of the dispersion of weak excitations~\cite{claude2023spectrum}, corresponding to the measurement of the analogue spacetime metric~\cite{Garay2000,Lahav2010}, including rotating configurations~\cite{Guerrero2025}. After a theoretical demonstration of the quantum analogue of the Penrose effect in a polariton condensate~\cite{solnyshkov2019quantum}, where the angular momentum of the black hole was changing in discrete quanta, some of us have shown that the wavevector-dependent losses in polariton condensates could make each vortex a center of convergent flow~\cite{solnyshkov2024towards}. Each quantum vortex can therefore become an analogue black hole, and a pair of vortices can exhibit mutual rotation, evolving towards a black hole merger. We have demonstrated that a merger of the static limits was indeed possible for a pair of vortices, but a merger of the horizons was remaining beyond reach so far.

In this work, we demonstrate that working with a sufficient number of quantum vortices allows one to study a complete analogue of the black hole merger, with the formation of a common horizon, whose properties are determined by simple geometric laws. We show an important difference between the true and the apparent horizons of the resulting analogue black hole. The properties of this black hole, such as its mass and angular momentum, determined by the integer (and relatively small) number of vortices, are quantized, which leads to interesting deviations from the perfectness of classical black holes.

\section{The model}

There were a lot of studies of analogue black holes in quantum fluids~\cite{Barcelo2005,Barcelo2018,almeida2023analogue}. We consider a particular configuration, where the quantum fluid providing the analogue spacetime is a Bose-Einstein condensate of cavity polaritons~\cite{kasprzak2006bose,HaiSon2015}. Cavity polaritons are the eigenstates of the two coupled oscillator model, describing the strong coupling of excitons and 2D cavity photons~\cite{weisbuch1992observation,Microcavities,carusotto2013quantum} characterized by a 2D in-plane wavevector $\bm{k}$. Polaritons combine the properties of the two coupled modes: the repulsive interactions of excitons and the  small effective mass of cavity photons. Both excitons and cavity photons are characterized by a finite lifetime, typically much shorter for the photon. The polariton lifetime depends of the exciton and photon fractions, which in turn depend on the in-plane wave vector $\bm{k}$. 

Polariton condensates can be created by non-resonant optical pumping: high-energy electrons and holes are injected into the system and form excitons, which thermalize and condense in the ground state ($k=0$) thanks to energy relaxation via exciton-phonon, exciton-electron and exciton-exciton interactions. Once a stationary condensate is formed, its state is characterized by the compensation of the losses by the inbound scattering, whereas all excited states exhibit non-compensated losses scaling as the square of their wave vector (that is, linear in energy of bare polariton modes).  This scaling of the losses has been predicted theoretically~\cite{Solnyshkov2014} and agrees with the experiments~\cite{Wertz2012}. In particular, it is at the heart of the recent experimental observation of KPZ scaling in polariton condensates~\cite{fontaine2022kardar}.

We describe the polariton condensate formed in such conditions using the modified Gross-Pitaevskii equation, also called hybrid Boltzmann-Gross-Pitaevskii equation~\cite{Solnyshkov2014}:
\begin{eqnarray}
i\hbar \frac{{\partial \psi }}{{\partial t}} &=&  - \left( {1 - i\Lambda } \right)\frac{{{\hbar ^2}}}{{2m}}\Delta \psi  + U \psi \nonumber\\
&+& \alpha {\left| \psi  \right|^2}\psi  + i(\gamma {e^{ - {n_{tot}}/{n_0}}}-\gamma_0)\psi
\label{gpeloss}
\end{eqnarray}
where $m$ is the polariton mass, $U$ is the confining potential (for example, a cylindrical mesa), $\alpha$ is the polariton-polariton interaction constant, $\Lambda$ is the constant characterizing the $k^2$-losses, $\gamma_0$ are the losses at $k=0$, and $\gamma$ is the prefactor of the gain term, saturated by the total density $n_{tot}$. We do not take into account the local reservoir saturation \cite{Wouters2007,Keeling2008,Wouters2010}, because we consider the case when the condensate dynamics is much faster than that of the reservoir.

To solve the modified Gross-Pitaevskii equation~\eqref{gpeloss} numerically, we used the 3rd-order Adams-Bashforth method \cite{Bashforth1883}, while the Laplacian was computed using the Fast Fourier Transform accelerated by the graphics processor unit~\cite{Reese2011}. The grid size used was $2048\times 2048$, with the spatial step  $h=0.25~\mu$m (the system size is therefore $L=512~\mu$m) and the time step $dt=2\times 10^{-16}$s, with the duration of the simulation times up to $1\times 10^{-9}$~s. 
The interaction constant taken was that of GaAs, experimentally measured~\cite{Ferrier2011} to be of the order of $5$~$\mu$eV$\cdot\mu$m$^2$, with the pumping parameters chosen in such a way that the  average interaction energy far from the cores of the vortices was $\alpha n\approx 1$~meV. Other parameters were also taken to correspond to GaAs cavities. The polaritonic  mass was $m=5\times 10^{-5}m_0$, where $m_0$ is the free electron mass. This gives a healing length $\xi=\hbar/\sqrt{2m\alpha n}\approx 1~\mu$m. The confinement potential $U(r)=U_0\Theta(r-R)$ (cylindrical mesa) had a radius of $R=246~\mu$m, and its height was $U_0=10$~meV ($\Theta$ is the Heaviside function). The  $k^2$-decay coefficient was taken as $\Lambda=0.1$ (unless specified otherwise).

In each simulation, the initial condition was taken as a constant density profile with a single vortex or with a superposition of vortices with a ring-like shape characterized by an initial separation distance. This is equivalent to seeding the condensation by a relatively strong probe with a spatial intensity and phase profiles that could be obtained for example by using a Spatial Light Modulator. The size of each vortex was chosen to correspond to the healing length expected from the stationary particle density of the homogeneous solution, in order to reduce the initial perturbations. 

\section{Single vortex as a black hole}
\begin{figure}
    \centering
    \includegraphics[width=1.0\linewidth]{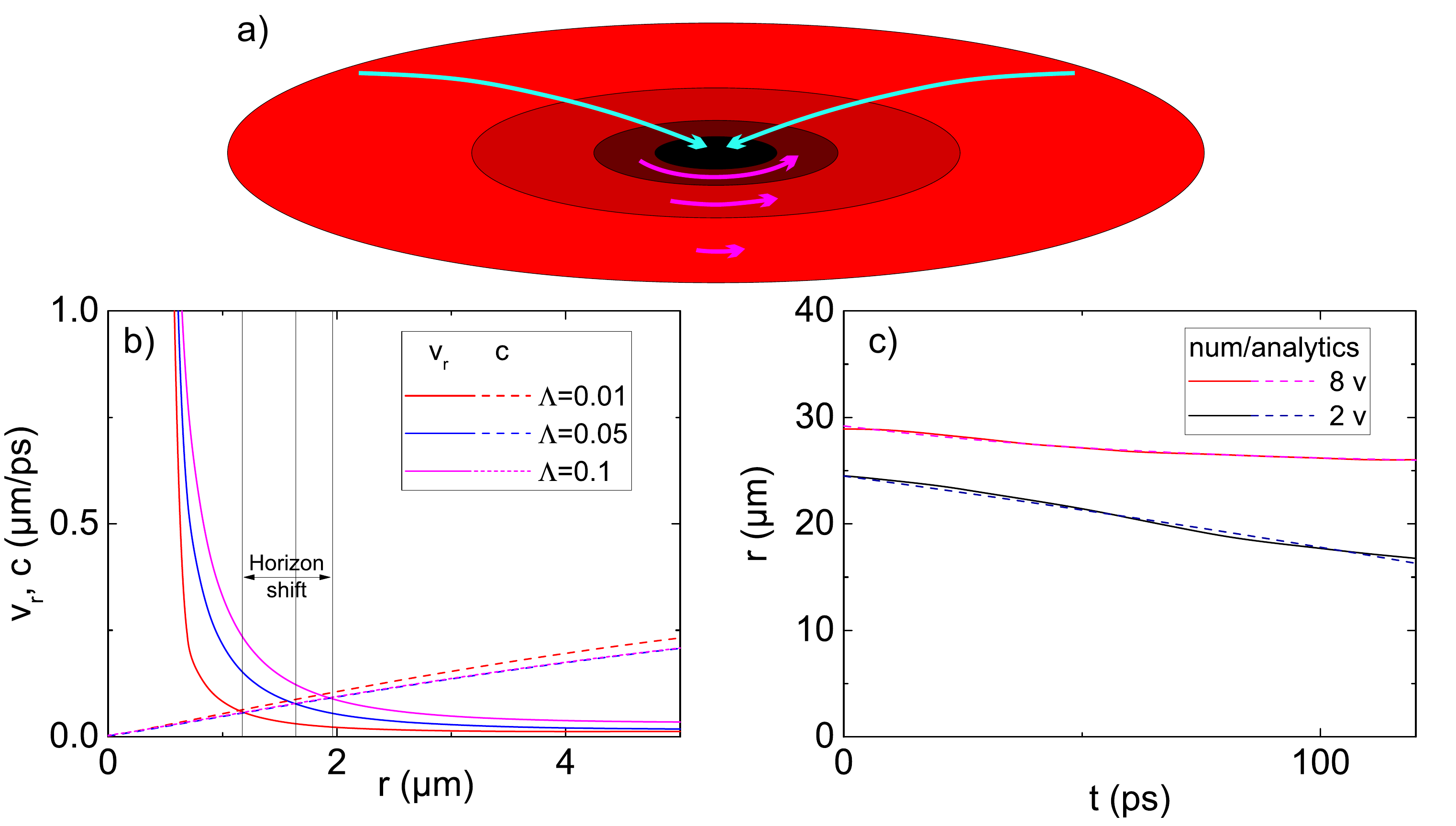}
    \caption{Polariton quantum vortices as analogue black holes. a) Scheme of a quantum vortex with local flow velocity components. b) Radial velocity $|v_r|$ and speed of sound $c$ as functions of distance $r$ from the vortex center for different damping $\Lambda$. Varying $\Lambda$ allows to change the radius of the horizon. c) The inspiral phase for 2 and 8 vortices: relative center of mass vortex coordinate $r$ as a function of time $t$. Solid lines -- numerical simulations, dashed lines -- analytical solution.}
    \label{fig1}
\end{figure}

We first briefly discuss a single vortex in a condensate with wavevector-dependent losses~\cite{solnyshkov2024towards}. Since the wave vector $-i\nabla\psi$ diverges in the center of the vortex because of the phase singularity, the losses also diverge. In a stationary configuration, they are compensated by a flow towards the center of the vortex. Each vortex therefore becomes a center of a convergent flow, as shown schematically in Fig.~\ref{fig1}a (cyan arrows -- radial component of the flow, magenta arrows -- azimuthal component, false color -- condensate density).

In order to be able to speak of a vortex as of an analogue black hole, we need to define the corresponding metric.
The metric of acoustic excitations in a general flow is totally determined \cite{Unruh1981} by the background stationary velocity $\textbf{v}=(\hbar/m)\nabla\arg\psi$ and the local speed of sound $c_s=\sqrt{\alpha |\psi|^2/m}$. We write it first in cartesian (2+1) coordinates:
\begin{equation}
    {g_{\mu \nu }} = \frac{{mn}}{{{c_s}}}\left( {\begin{array}{*{20}{c}}
  { - \left( {c_s^2 - {v^2}} \right)}&{ - {v_x}}&{ - {v_y}} \\ 
  { - {v_x}}&1&0 \\ 
  { - {v_y}}&0&1 
\end{array}} \right)
\label{gmetric}
\end{equation}
In a radially-symmetric flow with radial $v_r$ and azimuthal $v_\phi$ components of velocity, the metric is given by~\cite{Berti2004,Visser2005}:
\begin{equation}
\label{Kcm}
g_{\mu\nu}=\frac{mn}{c_s}
\begin{pmatrix} 
-(c_s^{2}-{v_\mathrm{tot}}^{2}) & 0 & -rv_\phi\\
0 & \bigg(1-\frac{v_r^2}{c_s^2}\bigg)^{-1} & 0 \\
-rv_\phi & 0 & r^2\\
 \end{pmatrix}
 \end{equation}
with $v_\mathrm{tot}$ the total velocity. To obtain the terms of this metric explicitly, one needs to use a solution for the wave function. Approximate analytical solution in presence of wavevector-dependent losses was presented in Ref.~\cite{solnyshkov2024towards}. 

Here, we focus on the possibilities for the control of the parameters of the black hole associated with the vortex, such as the radius of the horizon, using numerical simulations. One of the most important control parameters is the energy relaxation efficiency $\Lambda$. It can be tuned experimentally via  directly accessible parameters such as exciton-photon detuning and temperature~\cite{Solnyshkov2014}. Figure~\ref{fig1}(b) shows how changing $\Lambda$ affects the speed of sound $c_s$ and the radial velocity $v_r$ for a single vortex, thus changing the position of the horizon defined by $|v_r|=c_s$. As could be expected, radial velocity is affected more than the speed of sound, because the higher losses need to be compensated by a stronger inwards flow. For the range of values of $\Lambda$ we used (from 0.01 to 0.1), the radius of the horizon increases by a factor 2, which is quite significant. In what follows, we use a fixed value $\Lambda=0.1$.

\section{Vortex attraction}

With each vortex becoming a center of a convergent flow, multiple vortices are able to exhibit mutual attraction, which can even overcome the repulsion of same-sign vortices induced by the centrifugal force. In Ref.~\cite{solnyshkov2024towards} we have found an analytical solution for the time dynamics of a pair of vortices, which can be written in terms of a special function $\mathrm{Ei}$ (exponential integral):
\begin{equation}
    t(r)=t_0-\xi^2 a^{-1}\mathrm{Ei}\left(2\log \frac{r}{\xi}\right)
    \label{zamsol}
\end{equation}
However, obtaining a common horizon induced by the collective flow of the vortices was found to be impossible for only 2 vortices in Ref.~\cite{solnyshkov2024towards}.
In the present work, we consider multiple-vortex configurations of a ring-like shape. In such a case, the vortex attraction leads to the decrease of the radius of the ring. We begin by demonstrating that the analytical solution~\eqref{zamsol} applies even to the multiple-vortex case with $r$ being the distance between a vortex and the center of mass of the "vortex molecule" (being at the center of the ring). Figure~\ref{fig1}(c) presents a comparison of numerically-calculated dependence of the distance between the vortices  on time $r(t)$ for the reference case of 2 vortices (black) and for a particular case of 8 vortices (red). The numerical curves (solid lines) are nicely fitted by the analytical solution~\eqref{zamsol} (dashed lines) in both cases, thereby confirming the validity of our description of the inspiral phase of analogue black hole merger based on convergent flows even in the multiple-vortex case. 

\section{Common horizon}
We now focus on the determination of the position of the horizon for a system of several vortices, hoping to observe the formation of a common horizon. While a simultaneous merger of multiple black holes is a very rare event in astrophysics and merger of completely identical black holes is surely highly improbable, such configurations have been considered theoretically in numerical relativity~\cite{diener2003new}. One of the important differences between the astrophysical black holes and our analogue framework, beside the dimensionality, is that the astrophysical black holes are composed of a huge number of particles. They are described classically, and do not have any features associated with individual particles composing them. In our analogue system, we always deal with a small number of identical vortices, each of them representing a black hole. Even when they create a collective horizon, the resulting structure can still be expected to retain the information on its elementary ingredients, contrary to the featureless astrophysical black holes. This is a first step towards "quantum" black holes, if not yet towards quantum gravity.

\begin{figure*}
    \centering
    \includegraphics[width=1.0\linewidth]{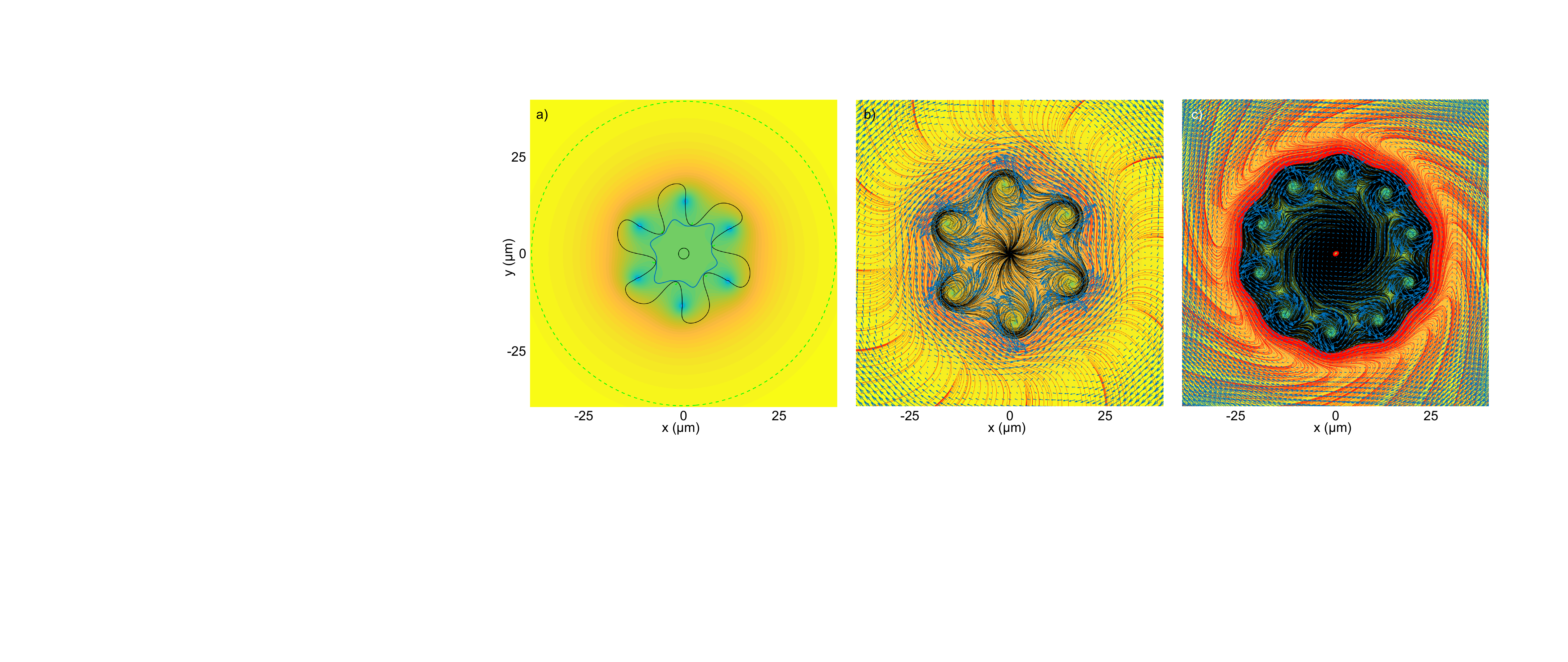}
    \caption{Formation of a common horizon. a) A merger of 6 quantum vortices.  Green dashed line -- static limit, black solid line -- $|v_r|=c$, blue line -- apparent horizon (Eq.~\eqref{horeq}). Vortex centers appear as blue dots. b,c) Simulation of possible trajectories on top of the background flow for 6 and 10 vortices. Arrows: background flow velocity in the rotating frame. Lines: possible trajectories (black -- infalling, red -- escaping). The true horizon appears as a boundary between red and black trajectories.  
    In all panels, false color represents the polariton density (log scale).}
    \label{fig2}
\end{figure*}

While it was possible to write the analogues of the Schwarzschild and the Kerr's metrics explicitly and analytically in the cylindrically-symmetric configurations, and thus determine the position of the horizon in both cases, the definition of the horizon is more complicated in a general case which is not cylindrically-symmetric and not stationary. A similar problem arises in numerical relativity, where the apparent horizon is often used for computational reasons instead of the true event horizon for non-stationary black hole configurations~\cite{centrella2010black}.
We use the following equation, which determines the apparent horizon from its qualitative definition as a surface (or contour in 2+1 dimensions) that no wave excitations can cross (in the acoustic limit). The equation reads:
\begin{equation}
    \vec{v}\cdot\vec{n}+c_s=0
    \label{horeq}
\end{equation}
where the $\vec{n}$ is the outward-pointing normal.
This equation is completely equivalent to the one used in numerical relativity for the determination of marginally trapped surfaces and apparent horizons~\cite{thornburg2003fast,campanelli2006accurate,pook2019existence}, where one requires future-pointing outgoing null geodesics (light beam trajectories) to have a zero expansion $\Theta=0$ with respect to the outward-pointing normal to the surface (no outward light beams). This is an implicit nonlocal equation, because $\vec{n}$ depends on the position of neighboring points of the horizon curve, and the scalar product depends not only on the local value of $\vec{v}$, but also on $\vec{n}$.
Despite the non-local character of the equation, the apparent horizon is a local property, valid at a given time. In general relativity, the difference between the true horizon and the apparent horizon can appear, for example, because of a temporal evolution of a black hole: the position of the event horizon can be determined only when the full time evolution is known~\cite{Alcubierre2001,thornburg2003fast}. A particle or a light beam which at some moment of time is outside an apparent horizon might be "caught" by the expansion of the horizon due to the increase of the black hole mass. It is actually impossible for this particle to escape to infinity, and therefore it must be contained inside the true horizon, while being outside the apparent one. In a non-cylindrically symmetric and/or non-stationary metric, a time-like trajectory can therefore be outside of the apparent horizon, but at the same time unable to reach the infinity. These globally bounded trajectories define the location of the true event horizon, as we will see later.

In order to determine the position of the horizon, we start by solving a simpler equation $v_r=-c$, with the radial direction determined with respect to the center of mass of the system of vortices. If a closed curve around the origin is obtained, it means that everywhere inside this curve, $v_r<-c_s$ (except a region close to $r=0$, always surrounded by a second curve with an opposite change of sign for symmetry reasons). It means that the whole curve is contained within the apparent horizon: at least, we can be sure that a signal cannot get outside of this circle. We use this curve as a starting approximation to reconstruct the apparent horizon numerically. This approach allows to find the apparent horizon much faster than by solving Eq.~\eqref{horeq} directly.

Figure~\ref{fig2}(a) presents an example of an accomplished black hole merger for 6 vortices. The dashed green curve is the static limit defined by $v_{tot}=c$. It is located at a large distance from the center of mass of the system and because of this it is essentially featureless (contrary to the 2-vortex case presented in~\cite{solnyshkov2024towards}). The black solid curve is the solution of $v_r=-c$, and the 6 points of this curve closest to the center of mass also belong to the apparent horizon, as discussed above. The apparent horizon given by \eqref{horeq} is represented  by a blue curve.

In order to determine the position of the true horizon in our analogue system, where the full time evolution is well known, we track an ensemble of possible trajectories of particles with different initial positions and with the maximal locally possible speed $c_s$, trying to escape from the black hole (Fig.~\ref{fig2}(b,c)). The simulation of these trajectories is performed in the rotating frame, where the positions of the vortices forming the black hole are stationary. The particles are carried with the known velocity of the background flow $\bm{v}_{fr}$ in this frame. Their velocity relative to the flow is directed at each step in the radial direction, so that their total velocity is: $\bm{v}_{part}=\bm{v}_{fr}+c_s\bm{u}_r$. We plot the resulting trajectories in Fig.~\ref{fig2}(b,c) for black holes composed of 6 and 10 vortices, respectively, with two colors: red for trajectories with $r(t_{max})>r(0)$ (that is, ultimately escaping) and black for trajectories with $r(t_{max})\le r(0)$ (infalling, bounded trajectories). The background flow velocity in the rotating frame $\bm{v}_{fr}$ is plotted with blue arrows. A clear boundary between red and black trajectories can be observed, indicating the position of the true horizon. All vortices are contained inside this horizon, demonstrating that the merger of the black holes has indeed occurred. We also note that the apparent horizon is contained inside the true horizon.

We note that the symmetry of the resulting horizon is not $C_{\infty}$ as for a relativistic Schwarzschild or Kerr black holes, but $C_n$, where $n$ is the number of vortices. Because of this, the metric of the resulting black hole cannot be written analytically. Formally, it is determined by the expression~\eqref{gmetric} in cartesian coordinates, but the functions $n$, $c_s$ and $\bm{v}$ that it contains must be determined numerically or experimentally.
The absence of cylindrical symmetry is an indication of the quantum nature of the constituents of the black hole. Some signatures of this quantum nature have already been studied theoretically in the quantum analogue of the Penrose effect~\cite{solnyshkov2019quantum}. The higher the number of the vortices, the smaller the deviation from $C_\infty$, so that for a black hole with a macroscopic (and very large) number of particles one can indeed recover the perfect symmetry predicted by general relativity~\cite{Werner1967}, which is a classical theory.

\begin{figure}
    \centering
    \includegraphics[width=1.0\linewidth]{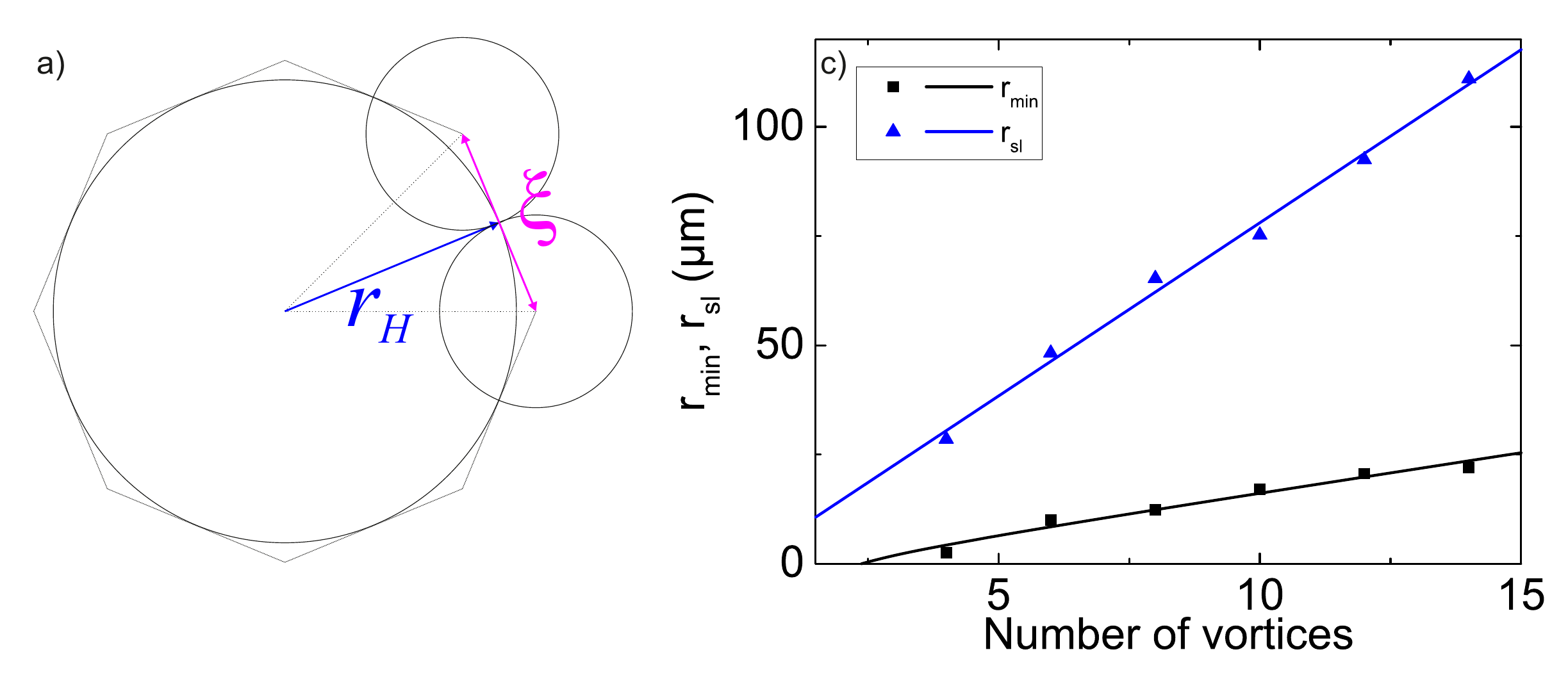}
    \caption{Multiple-vortex black hole. a) Illustration for the link between the number of vortices and the radius of the horizon. b) Fine structure of the horizon explaining its modulation. c) Minimal radius of the apparent horizon and the static limit as a function of number of vortices. Black dots -- numerical simulations, solid lines -- analytical fits.    
    }
    \label{fig3}
\end{figure}

\section{Radius of the horizon}

The radius of the horizon is an important property of black holes~\cite{Taylor}. It scales linearly with the black hole mass, and it also depends on the angular momentum of the black hole. It is therefore important to analyze the behavior of the radius of the common horizon of an analogue black hole.
In this section, we study this behavior as a function of number of vortices. In order to obtain an analytical estimate of the radius of the horizon, we represent the system at the moment of the merger as a circular vortex chain, as studied experimentally in Ref.~\cite{Boulier2015}, for example. The minimal distance between the vortices in a chain is given by the healing length $\xi$, which allows to find the radius of the horizon approximately as the radius of a circle inscribed in a polygon with the number of sides equal to the number of vortices $n_v$ and the length of each side given by $\xi$: 
\begin{equation}
    r_h\approx \frac{\xi}{2\tan\left(\pi/n_v\right)}
    \label{rhest}
\end{equation}
This is shown schematically in Fig.~\ref{fig3}(a), with small circles of the size of the healing length representing the vortices, and the big circle representing the estimate for $r_h$.

In Fig.~\ref{fig3}(c) we show the distance between the center of mass and the closest point of the horizon (which can be called its smallest radius) obtained from numerical simulations as a function of the number of vortices with black squares, while the analytical solution~\eqref{rhest} is shown with a black solid line. Similarly, we show the radius of the static limit with blue triangles, and the analytical prediction based on the expression from~\cite{solnyshkov2019quantum} as a blue solid line ($r_{sl}\sim \xi n_v$ for sufficiently large $n_v$). The good agreement between the numerical simulations and the analytical estimates confirms the validity of our interpretation. 
For a sufficiently large number of vortices $n_v$, one obtains $r_h\sim n$, which means that the behavior of the analogue black hole is similar to that of astrophysical black holes: the radius of its horizon grows linearly with the number of particles composing it (that is, linearly with its mass).

\section{Discussion and conclusions}

In our numerical simulations, a common horizon is not formed for a number of vortices smaller than 4. This can also be understood from geometric reasons, in agreement with Eq.~\eqref{rhest}. We did not consider higher numbers of vortices than $n_v=14$, taking into account the realistic parameters of the SLMs used in current experiments. Working with a relatively small number of vortices represents a limit opposite to the one studied experimentally in liquid helium, where a giant vortex with $n_v\sim 10^4$ has been created recently~\cite{vsvanvcara2024rotating}. According to Eq.~\eqref{rhest} (in agreement with numerical simulations), the relative amplitude of the modulation of the apparent horizon decreases with the number of vortices, with the system tending towards the classical featureless black hole limit.

To conclude, we have demonstrated the formation of a common event horizon for a system of multiple vortices as an accomplishment of an analogue black hole merger after the inspiral phase. We compute the shape of this horizon and determine its parameters as a function of the number of vortices. These findings open large opportunities for the field of analogue gravity.

\begin{acknowledgments}
Our work was supported by the ANR program "Investissements d'Avenir" through the IDEX-ISITE initiative 16-IDEX-0001 (CAP 20-25), the ANR project MoirePlusPlus (ANR-23-CE09-0033), and the ANR project HAWQ (ANR-25-CE47-7323).
\end{acknowledgments}

\bibliography{biblio}

\end{document}